\documentclass[%
reprint,
amsmath,amssymb,
aps,
]{revtex4-2}

\usepackage{lipsum}
\usepackage{bm}
\usepackage{amssymb}
\usepackage{amstext}
\usepackage{amsmath}
\usepackage{mathrsfs}
\usepackage{graphicx}
\usepackage{color}
\usepackage{amsthm}
\usepackage{floatrow}
\usepackage{array}

\usepackage{listings}
\usepackage{booktabs}
\usepackage{changepage}
\usepackage{hyperref}

\begin{document}
	\preprint{APS/123-QED}
	\title{Fast renormalizing the structures and dynamics of ultra-large systems via random renormalization group}
	\thanks{Correspondence should be addressed to Yizhou Xu and Pei Sun.}%

 \author{Yang Tian}
	\email{tyanyang04@gmail.com \& tiany20@mails.tsinghua.edu.cn}
	\altaffiliation[]{Department of Psychology \& Tsinghua Laboratory of Brain and Intelligence, Tsinghua University, Beijing, 100084, China.}

	\author{Yizhou Xu}%
\email{xuyz23@mails.tsinghua.edu.cn}
	\altaffiliation[]{Department of Mathematical Sciences, Tsinghua University, Beijing, 100084, China.}
	
	\author{Pei Sun}%
	\email{peisun@tsinghua.edu.cn}
	\altaffiliation[]{Department of Psychology \& Tsinghua Laboratory of Brain and Intelligence, Tsinghua University, Beijing, 100084, China.}

	
	
	
\begin{abstract}
Criticality and symmetry, studied by the renormalization groups, lie at the heart of modern physics theories of matters and complex systems. However, surveying these properties with massive experimental data is bottlenecked by the intolerable costs of computing renormalization groups on real systems. Here, we develop a time- and memory-efficient framework, termed as the random renormalization group, for renormalizing ultra-large systems (e.g., with millions of units) within minutes. This framework is based on random projections, hashing techniques, and kernel representations, which support the renormalization governed by linear and non-linear correlations. For system structures, it exploits the correlations among local topology in kernel spaces to unfold the connectivity of units, identify intrinsic system scales, and verify the existences of symmetries under scale transformation. For system dynamics, it renormalizes units into correlated clusters to analyze scaling behaviours, validate scaling relations, and investigate potential criticality. Benefiting from hashing-function-based designs, our framework significantly reduces computational complexity compared with classic renormalization groups, realizing a single-step acceleration of two orders of magnitude. Meanwhile, the efficient representation of different kinds of correlations in kernel spaces realized by random projections ensures the capacity of our framework to capture diverse unit relations. As shown by our experiments, the random renormalization group helps identify non-equilibrium phase transitions, criticality, and symmetry in diverse large-scale genetic, neural, material, social, and cosmological systems.

\end{abstract}
\maketitle
We live in an era where physics faces challenges arising from the massive data of the world. The costs of computing physics theories become increasingly intolerable as target system size grows \cite{hendrickson2009computational,press2007numerical}. The applicability of analytic formalism is frequently limited by the vague details of real systems \cite{kutz2013data,kirchdoerfer2016data}. These challenges give rise to an opportunity to integrate physics with advanced computational science \cite{post2005computational}. Recently, this pursuit has achieved substantial progress in condensed matter physics \cite{ohno2018computational,bedolla2020machine,schmidt2019recent,schleder2019dft}, quantum chemistry \cite{mcardle2020quantum,bauer2020quantum,cao2019quantum}, and fluid mechanics \cite{ferziger2019computational,brunton2020machine,anderson2020computational}.

However, there remain numerous physics theories lacking ideal computational frameworks, among which, renormalization group (RG) \cite{goldenfeld2018lectures,efrati2014real} is an important one. A renormalization group is a fundamental tool for studying criticality and symmetry in system structure (i.e., the connectivity of units) and dynamics (i.e., the dynamic behaviours of units across time), which defines scale transformation of the system utilizing the correlations among system units \cite{pelissetto2002critical,jona2001renormalization}. Its universal applicability to different systems puts non-negligible challenges for computational designs. First, the computational complexity of an RG pipeline should be as low as possible since scaling analysis usually requires large system sizes \cite{kadanoff2000statistical,reichl2016modern}. Second, the numerical implementation and optimization of RGs cannot be domain-specific to reduce universality. Third, an ideal RG framework should be capable of characterizing different kinds of unit correlations, irrespective of whether they are  linear or non-linear. Fourth, the application on real data may require an RG framework to reduce the dependence on the \emph{a priori} knowledge about the mechanisms underlying data generation (e.g., system Hamiltonian or internal organization) since such information can be unavailable \cite{nicoletti2020scaling}. Considering these challenges, one would find existing computational realizations of RGs imperfect. Among these works, although Monte-Carlo-based \cite{ron2017surprising,troster2015fourier,wu2019determination,ron2002inverse,wu2017variational,wu2020continuous}, discriminative-model-based \cite{bachtis2021adding,di2022deep}, and generative-model-based RGs \cite{koch2018mutual,li2018neural,chung2021neural,hu2022rg} are effective in estimating coarse-grained configurations and system parameters, they demand the \emph{a priori} knowledge about target system as inputs, relay on specialized optimization before application, and lack the generalization capacity to new systems unless extra optimization is supported. Compared with these machine-learning-aid frameworks, the optimization-free designs of phenomenological RGs \cite{villegas2023laplacian,loures2023laplacian,wei2019sampling,song2005self,song2007calculate,garcia2018multiscale,zheng2023geometric,gfeller2007spectral,nicoletti2020scaling,bradde2017pca,redman2020renormalization} are more favorable in reducing the reliance on \emph{a priori} knowledge and training. In general, these RGs apply pre-defined statistical rules to iteratively renormalize system structures or dynamics and verify specific scaling behaviours derived beforehand. Their main limitations are the inevitable trade-offs among high time complexities, heavy memory usages, and the sufficient capacities to accurately describe complex correlations. 

    \begin{figure*}[!t]
\includegraphics[width=1\columnwidth]{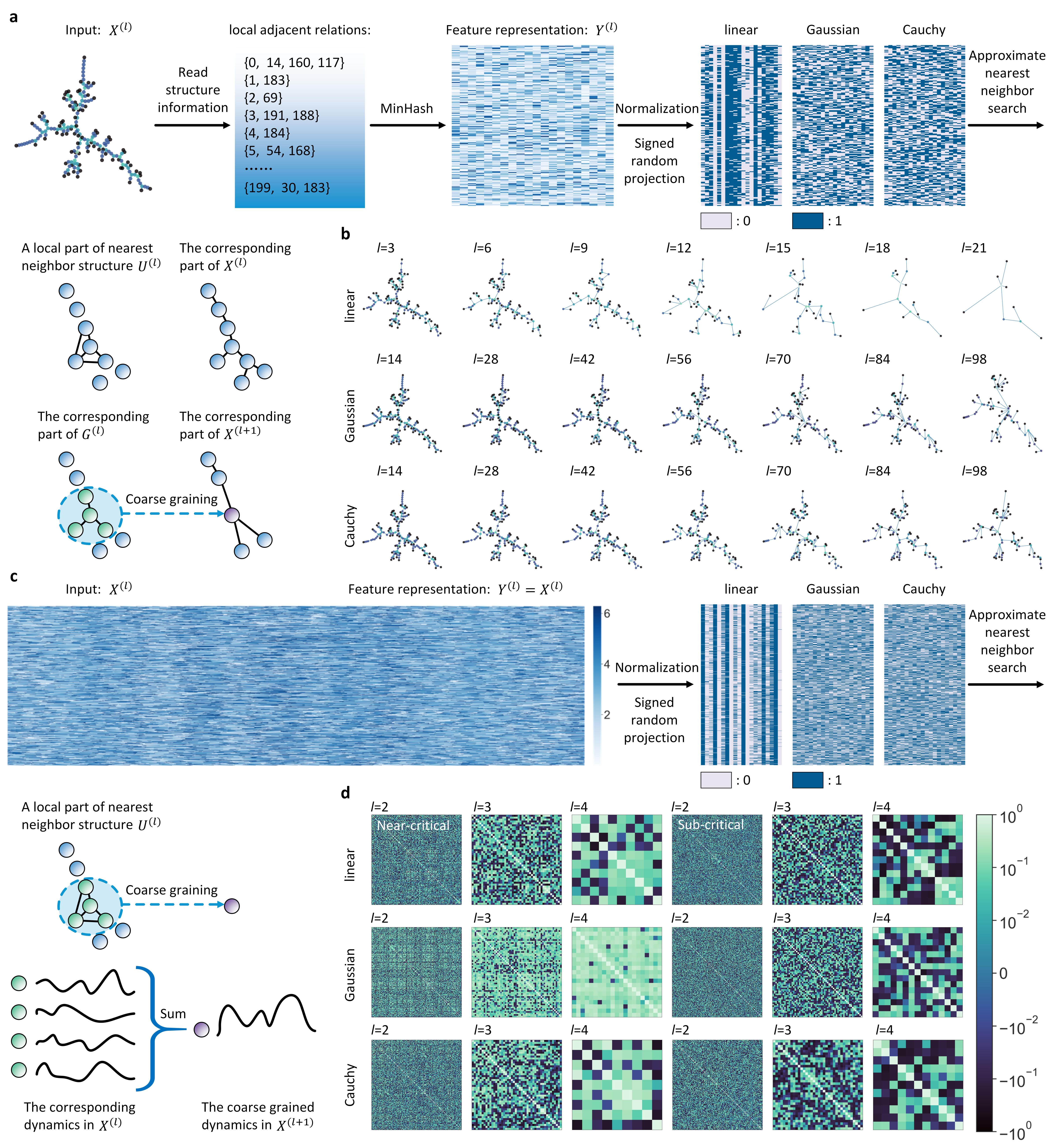}
\caption{\label{G0} Conceptual illustrations of the RRG framework. \textbf{a}, The pipeline of structure renormalization in the RRG. \textbf{b}, The structure renormalization flows of a random tree in different kernel spaces. \textbf{c}, Key steps of dynamics renormalization in the RRG. \textbf{d}, The dynamics renormalization flows of a system of $1000$ Kuramoto oscillators \cite{acebron2005kuramoto} under critical (i.e., coupling strength approximates to $1$) and sub-critical (i.e., coupling strength is $0.1$) conditions. Correlation matrices are used to represent unit relations and indicate system evolution during renormalization.} 
\end{figure*}

    Here, we develop a time- and memory-efficient framework, named as the random renormalization group (RRG), to resolve the trade-offs faced by optimization-free RGs. Established on random projections \cite{mahoney2011randomized} and hashing techniques \cite{cai2019revisit}, the RRG reduces computational complexity significantly (see Fig. \ref{G4}a) and supports a fast analysis of ultra-large systems using personal computers (see Figs. \ref{G4}b-c for accelerations compared with other RGs). Different from the classic approaches proposed only for structure \cite{villegas2023laplacian,loures2023laplacian,wei2019sampling,song2005self,song2007calculate,garcia2018multiscale,zheng2023geometric,gfeller2007spectral} or dynamics renormalization \cite{nicoletti2020scaling,bradde2017pca,redman2020renormalization}, the RRG can be applied to study criticality and symmetry in both system structures and dynamics. Its kernel space designs \cite{hofmann2008kernel,shawe2004kernel} enhance the expression ability of different kinds of correlations among system units to meet diverse research demands. As applications, the RRG is used to discover scale-invariant structures and critical dynamics in diverse large-scale genetic, neural, material, social, and cosmological systems. We release the code implementation of the RRG as an open-source toolbox in Ref. \cite{RRGrelease}.

    \section{Random renormalization group}
    In general, the key idea of the renormalization group is to define correlations among units, integrate out short-range correlations by coarse-graining, and select long-range correlations to describe the next scale \cite{pelissetto2002critical,jona2001renormalization}. In the RRG, this idea is realized by an organic combination of random projections, hashing techniques, and kernel representations. Given a system $X=\left(X_{1},\ldots,X_{N}\right)$ formed by $N$ units, we set it as the input of the RRG, i.e., $X^{\left(1\right)}=X$. For structure renormalization, system $X$ should describe the adjacent relations of $N$ units (e.g., being a network). For dynamics renormalization, each unit should have a time series of its activities. In each $l$-th iteration ($l\geq 1$), the RRG implements the following procedures:
    \begin{itemize}
        \item[(1) ] Define a feature representation, $Y^{\left(l\right)}$, of $X^{\left(l\right)}$ such that each unit $X^{\left(l\right)}_{i}$ has a feature vector $Y^{\left(l\right)}_{i}$ to represent its properties. Normalize each $Y^{\left(l\right)}_{i}$ following certain criteria for preparation.
        \item[(2) ] Apply a signed random projection to hash $Y^{\left(l\right)}$ as a binary representation $Z^{\left(l\right)}$ such that the Hamming distance between $Z^{\left(l\right)}_{i}$ and $Z^{\left(l\right)}_{j}$ approximates the correlation distance (i.e., a distance changes inversely with correlation) between $Y^{\left(l\right)}_{i}$ and $Y^{\left(l\right)}_{j}$ in a specific kernel space. 
        \item[(3) ] Realize an approximate nearest neighbor search on $Z^{\left(l\right)}$ to find the nearest neighbor of each unit $X^{\left(l\right)}_{i}$. All pairs of nearest neighbor relations are included in space $U^{\left(l\right)}$.
        \item[(4) ] Define a null network, $G^{\left(l\right)}$, of all units (i.e., a network without edge). For dynamics renormalization, edges are added between all pairs of nearest neighbors in $U^{\left(l\right)}$. For structure renormalization, an edge is added between a pair of nearest neighbors in $U^{\left(l\right)}$ only if they are adjacent in $X^{\left(l\right)}$ too. After edge adding, every connected cluster, $C^{\left(l\right)}_{k}$, of network $G^{\left(l\right)}$ includes the units sharing strong correlations.
        \item[(5) ] Coarse grain the units in each connected cluster $C^{\left(l\right)}_{k}$ into a macro-unit $X^{\left(l+1\right)}_{k}$. For dynamics renormalization, macro-unit $X^{\left(l+1\right)}_{k}$ is defined with a summed feature vector $Y^{\left(l+1\right)}_{k}=\sum_{i\in I^{\left(l\right)}_{k}}  Y^{\left(l\right)}_{i}$, where $I^{\left(l\right)}_{k}$ denotes the index set of all units in $C^{\left(l\right)}_{k}$. For structure renormalization, two macro-units are connected in $X^{\left(l+1\right)}$ if the units aggregated into them share at least one edge in $X^{\left(l\right)}$.
    \end{itemize}
    
    To appropriately describe system properties in step (1), we define $Y^{\left(l\right)}_{i}$ as the time series of unit $X^{\left(l\right)}_{i}$ to reflect dynamics (see Fig. \ref{G0}c for examples). As for structures, we define $W^{\left(l\right)}_{i}$ as the set where $X^{\left(l\right)}_{i}$ and all its adjacent units are included (see Fig. \ref{G0}a for instances). The Jaccard distance between $W^{\left(l\right)}_{i}$ and $W^{\left(l\right)}_{j}$ reflects the difference between $X^{\left(l\right)}_{i}$ and $X^{\left(l\right)}_{j}$ in terms of local topology (i.e., local adjacent relations). To avoid the potential difficulties of subsequent processing caused by the different set sizes of $W^{\left(l\right)}_{i}$ and $W^{\left(l\right)}_{j}$ and to reduce computational complexity, we apply the MinHash method \cite{broder1997resemblance,broder1998min} to hash each set $W^{\left(l\right)}_{i}$ as a feature vector, $Y^{\left(l\right)}_{i}$, of a given dimension (see Methods, Fig. \ref{G0}a, and Fig. \ref{G0}c). The normalized XOR distance between $Y^{\left(l\right)}_{i}$ and $Y^{\left(l\right)}_{j}$ approximates the concerned Jaccard distance (see Methods).

          \begin{figure*}
\includegraphics[width=1\columnwidth]{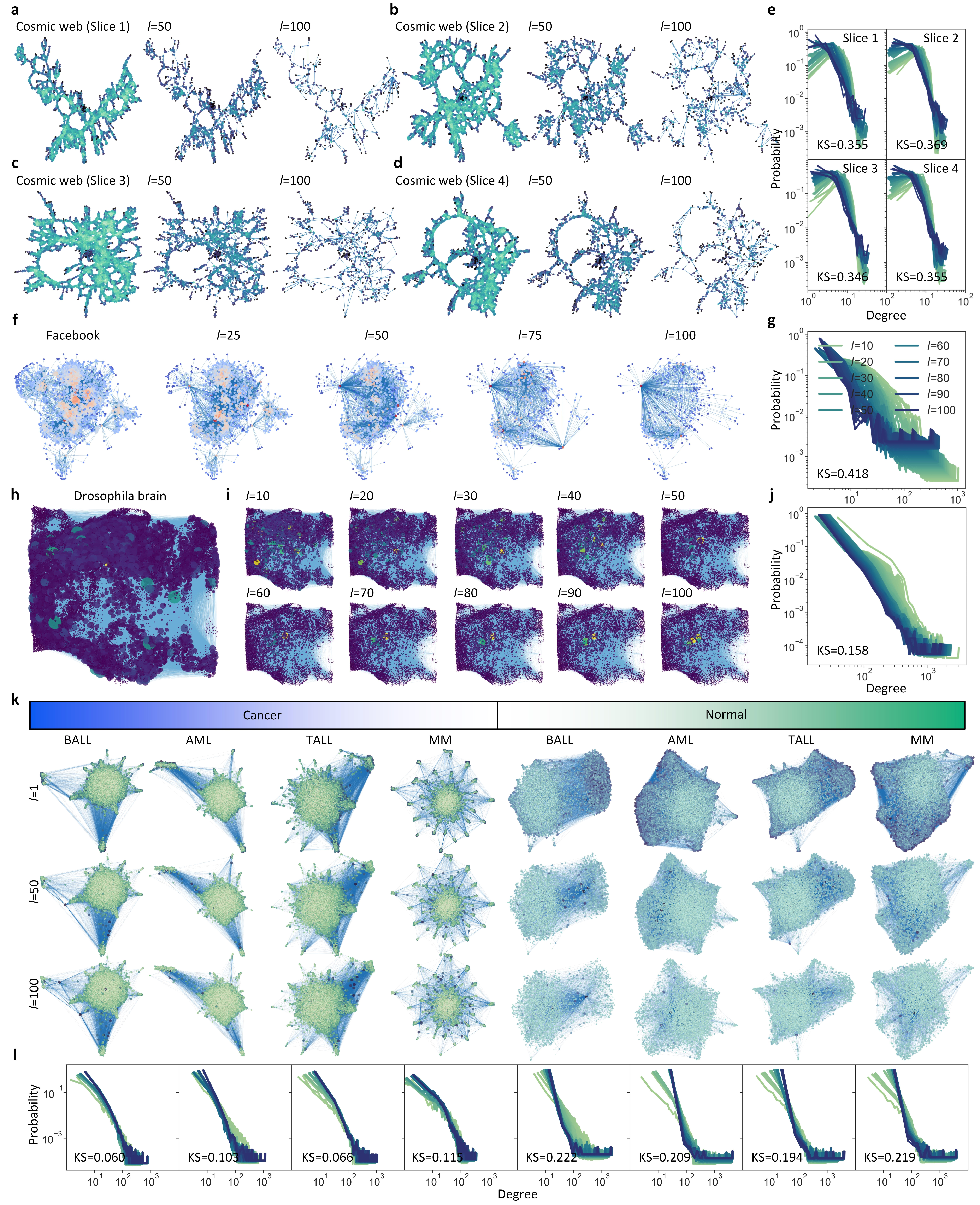}
\begin{adjustwidth}{-0cm}{}
    \caption[]{\label{G1}Structure renormalization on the real data. Sub-figures respectively show the renormalization flows, degree distributions, and Kolmogorov-Smirnov (KS) statistics of comic webs (\textbf{a-e}), Facebook network (\textbf{f-g}), Drosophila brain network (\textbf{h-j}), and gene co-expression networks (\textbf{k}).}
     \end{adjustwidth}

\end{figure*}

    To ensure the capacity to model different kinds of correlations in step (2), the RRG supports to apply the signed random hyperplane projection \cite{vempala2005random}, the signed random Fourier feature \cite{li2022signrff,rahimi2007random}, and the signed Cauchy projection \cite{li2013sign} to map $Y^{\left(l\right)}$ to $Z^{\left(l\right)}$ (see Methods, Fig. \ref{G0}a, and Fig. \ref{G0}c). Each kind of projection is equipped with a specific normalization procedure in step (1) (see Methods). Based on the first two kinds of projections, the collision probabilities between two hashed representations, $Z^{\left(l\right)}_{i}$ and $Z^{\left(l\right)}_{j}$, change inversely with the cosine similarity between $Y^{\left(l\right)}_{i}$ and $Y^{\left(l\right)}_{j}$ in the linear \cite{vempala2005random} and Gaussian kernel spaces \cite{li2022signrff}. Built on the third kind of projection, the collision probability varies oppositely with the $\chi^{2}$ similarity between $Y^{\left(l\right)}_{i}$ and $Y^{\left(l\right)}_{j}$ (i.e., a kind of correlation in the $\chi^{2}$ kernel space) \cite{li2013sign}. Therefore, these linear and non-linear correlations can all be efficiently approximated by the Hamming distance after hashing. In applications, the hashing based on signed random hyperplane projection serves as a robust and general approach while the signed random Fourier feature is better at characterizing units with strong correlations \cite{li2022signrff}. The hashing with a signed Cauchy projection is suitable for the cases where correlations are defined between the distributions of feature vector components \cite{li2013sign}. 

    The processing in steps (1-2) enables us to construct an approximate nearest neighbor search (see Methods, Fig. \ref{G0}a, and Fig. \ref{G0}c) \cite{liu2004investigation,li2019approximate}, which avoids calculating the whole correlation matrix among units during selecting short-range correlations (e.g., see opposite situations in the phenomenological RG \cite{nicoletti2020scaling}) and realizes significant acceleration and memory-saving in large-scale systems.

     Established on these designs, the RRG can efficiently unfold system structures and dynamics (see Fig. \ref{G0}b and Fig. \ref{G0}d for instances). One can see tutorials about the RRG code implementation in supplementary materials. It measures the correlations among local region topology properties or unit activities across time to progressively integrate out short-range correlations. The selected long-range correlations by the RRG reveal the defining organizational patterns or intrinsic dynamics of system units. The symmetry under the RRG transformation reflects the scale-invariance of concerned structures or dynamics. Below, we derive a series of macroscopic observables and scaling behaviours to analyze this kind of scale-invariance.

    \section{Scaling behaviours}
    
    To reflect the evolution of system structures during renormalization, we use the degree distribution of units as a macroscopic observable. Certainly, other macroscopic observables, such as the Laplacian eigenvalue spectrum \cite{villegas2023laplacian}, are acceptable as well. Here the degree distribution is selected due to its lower computational costs. 
    
    The potential scale-invariance of structures manifests as the phenomenon with a fixed degree distribution under the RRG transformation (i.e., the renormalized degree distribution does not departure from its original form significantly). These departures can be measured by a two-sided Kolmogorov-Smirnov test \cite{simard2011computing,berger2014kolmogorov} between the degree distribution and its coarse-grained counterpart (see Methods). In the application where the RRG runs $T$ iterations, we calculate the Kolmogorov-Smirnov statistic between the degree distributions derived on $X^{\left(l\right)}$ and $X^{\left(1\right)}$ for each $l\in\left[2, T\right]$ and average the statistic across  iterations. Given a short renormalization flow (e.g., when $T\leq 10$), we suggest $0.05$ as the threshold of the averaged statistic for determining whether the degree distribution is fixed. When the renormalization flow is long (e.g., when $T\geq 50$), an empirical choice of the threshold is suggested as $0.1$ or $0.15$. Here we do not use a strictly zero threshold because the RRG transformation reduces system size. As the renormalization continues, a finite real system, irrespective of how large it is, will inevitably become small and challenging for probability distribution estimation. Even in a case where the degrees on a coarse-grained scale are identically distributed with those on a fine-grained scale (e.g., all of them follow a distribution $P\left(\cdot\right)$), the small sample set of degrees on a coarse-grained scale may not ensure an observed degree distribution perfectly converged to $P\left(\cdot\right)$.

    For system dynamics, we also use a macroscopic observable to describe its variation during renormalization. In most applications, the probability distribution of normalized unit activities is a practical choice (see Methods). When the system exhibits critical dynamics near phase transitions, the distribution of normalized unit activities is expected to have a non-trivial (i.e., non-Gaussian) fix point. Under non-critical conditions (e.g., units are weakly correlated), the distribution either converges to a Gaussian fix point implied by the central limit theorem \cite{renyi2007probability} or lacks a fix point.

    In addition to the macroscopic observable, we can also analyze the scaling of system dynamics. Generalized from the phenomenological RG \cite{nicoletti2020scaling}, the RRG enables us to analyze diverse scaling features of system dynamics, including those about variance, free energy, covariance matrix eigenvalue spectrum, and correlation function (see Methods). In general, the scaling exponent, $\alpha\in\left[1,2\right]$, of the variance quantifies whether unit activities are strictly independent of each other ($\alpha=1$) or perfectly correlated with each other ($\alpha=2$). Meanwhile, the silence probability of the unit (i.e., the probability for dynamics to vanish) has a decay speed measured by the scaling exponent, $\beta\in\left[0,1\right]$, of free energy. Unit activities become increasingly independent when $\beta$ increases to $1$. Moreover, as a consequence of correlation function decay, the covariance matrix eigenvalue spectrum has a rank decay determined by scaling exponent $\mu$. Finally, the existence of dynamic scaling in unit activities implies a scaling behaviour of the characteristic time scale (i.e., also known as the correlation time, $\tau_{c}$) defined by  scaling exponent $\theta$. All these exponents are calculated according to the RRG flow (see Methods). Specifically, we denote the size, $K$, of a macro-unit in $X^{\left(l\right)}$ as the number of initial units (i.e., the original units in $X^{\left(1\right)}$) aggregated into it. Different from the phenomenological RG that requires a strict power form of the mean size, i.e., $K=2^{l-1}$ \cite{nicoletti2020scaling}, there is no constraint on $K$ in the RRG. Instead, the RRG focuses on the mean size of macro-units, $\langle K\rangle$, during scaling analysis. As the RRG iterates, the mean variance of unit activities (i.e., averaged across all units in each iteration) is expected to scale according to $\operatorname{Var}\left(\langle K\rangle\right)\propto \langle K\rangle^{\alpha}$. Similarly, the free energy satisfies $F\left(\langle K\rangle\right)\propto \langle K\rangle^{\beta}$ and the $r$-th largest eigenvalue of the covariance matrix exhibits a rank scaling $\lambda_{r}\propto \left(\langle K\rangle/r\right)^{\mu}$. Furthermore, the dynamic scaling of $\tau_{c}$ is expected as $\tau_{c}\left(\langle K\rangle\right)\propto \langle K\rangle^{\theta}$, where $\tau_{c}$ defines the exponential decay of the mean autocorrelation function, $C\left(t,\langle K\rangle\right)=\exp\left(-t/\tau_{c}\left(\langle K\rangle\right)\right)$ (averaged across all units in each iteration). The dynamic scaling also implies a universal collapse of $C\left(t,\langle K\rangle\right)=\exp\left(-t/\tau_{c}\left(\langle K\rangle\right)\right)$ after time re-scaling, $t\rightarrow t/\tau_{c}\left(K\right)$ (i.e., after re-scaling, the mean autocorrelation function derived on different iterations should follow a similar curve).

    As demonstrations, we implement the RRG on various real systems to realize the analysis presented above. One can also see supplementary materials for validating the RRG and the defined macroscopic observable in classifying random networks in terms of scale-invariance.

\section{Structure renormalization}
To prove the universal applicability of the RRG, we select real data from distinct scientific fields (see Methods for data processing). For convenience, the RRG runs 100 iterations, the signed random hyperplane projection is used in the RRG, and each binary representation, $Z^{\left(l\right)}_{i}$, has a dimension of $20$.

  \begin{figure*}
\includegraphics[width=1\columnwidth]{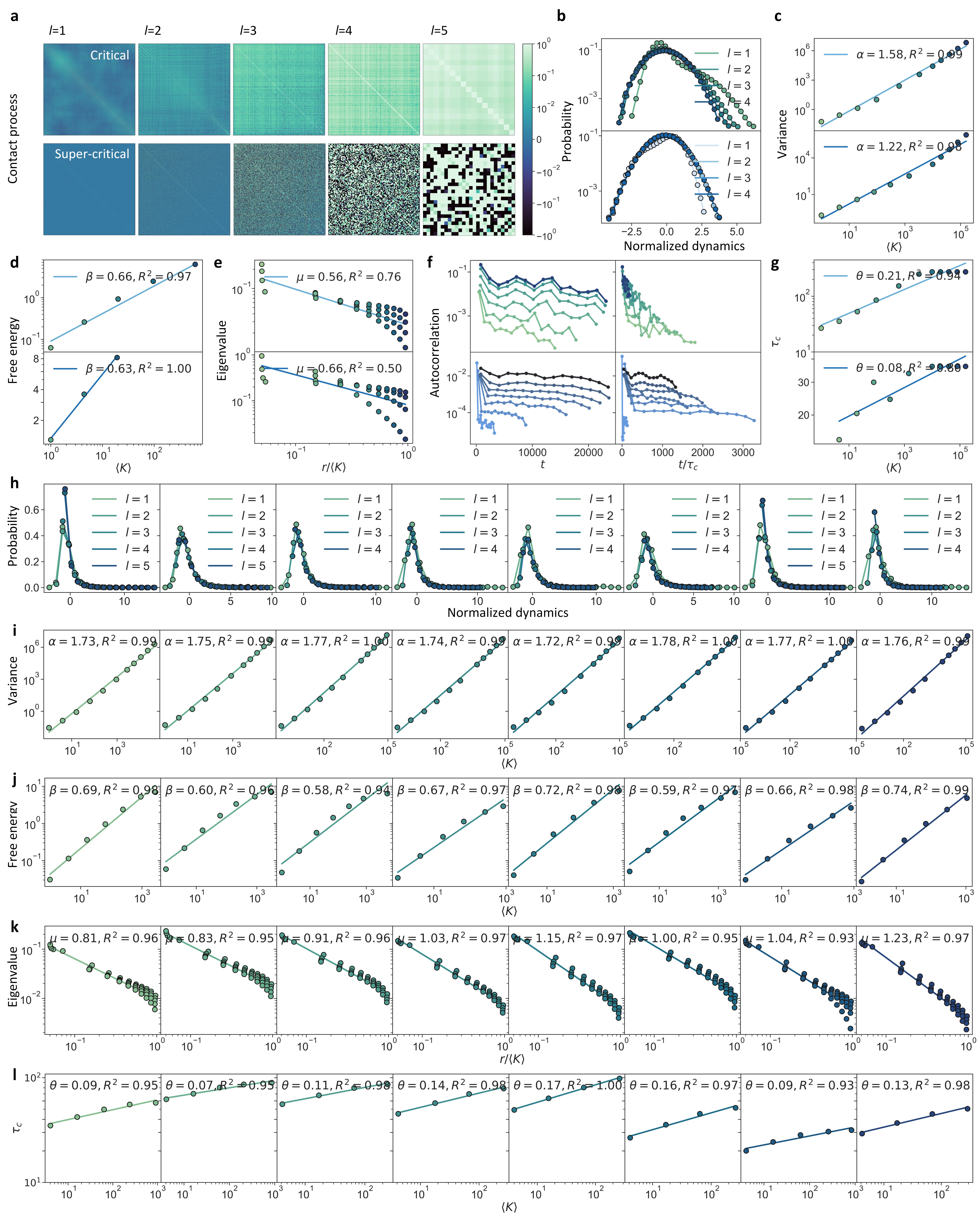}
\end{figure*}
\begin{figure*}
\begin{adjustwidth}{-0cm}{}
    \caption[]{\label{G2}Dynamics renormalization on the real data. \textbf{a}, The renormalization flows of a contact process with $10000$ units under critical and super-critical conditions. \textbf{b}, The probability distributions of normalized dynamics during renormalization. \textbf{c-e} and \textbf{g}, The scaling behaviours of the contact process characterized by exponents $\alpha$, $\beta$, $\mu$, and $\theta$. \textbf{f}, The existence and vanishing of universal collapse in the autocorrelation functions of critical and super-critical contact processes. \textbf{h-i}, The normalized dynamics distributions and scaling behaviours of eight zebrafish larva brains, where each column corresponds to the whole-brain dynamics data of one brain.}
     \end{adjustwidth}
\end{figure*}

For cosmology, we choose the cosmic web data derived from a high resolution cosmological magnetohydrodynamics simulation \cite{CosmologicalDataBase,vazza2020quantitative}. The cosmic volume is subdivided into four slices, in which the filtered galaxies are treated as units, and the filaments of ordinary and dark matters connecting between galaxy clusters serve as edges (see Methods). The cosmic web in each slice is inputted to the RRG as $X^{\left(1\right)}$ in Figs. \ref{G1}a-d. In Fig. \ref{G1}e, significant deviations from the initial degree distributions can be seen in all slices, suggesting the lack of scale-invariance in these cosmic webs.

For social science, we use Facebook data where units denote users and edges describe friendships (Fig. \ref{G1}f) \cite{FacebookDataBase}. Similar to cosmic webs, the friendship network is not scale-invariant under the RRG transformation in Fig. \ref{G1}g.

 \begin{figure*}[!t]
\includegraphics[width=1\columnwidth]{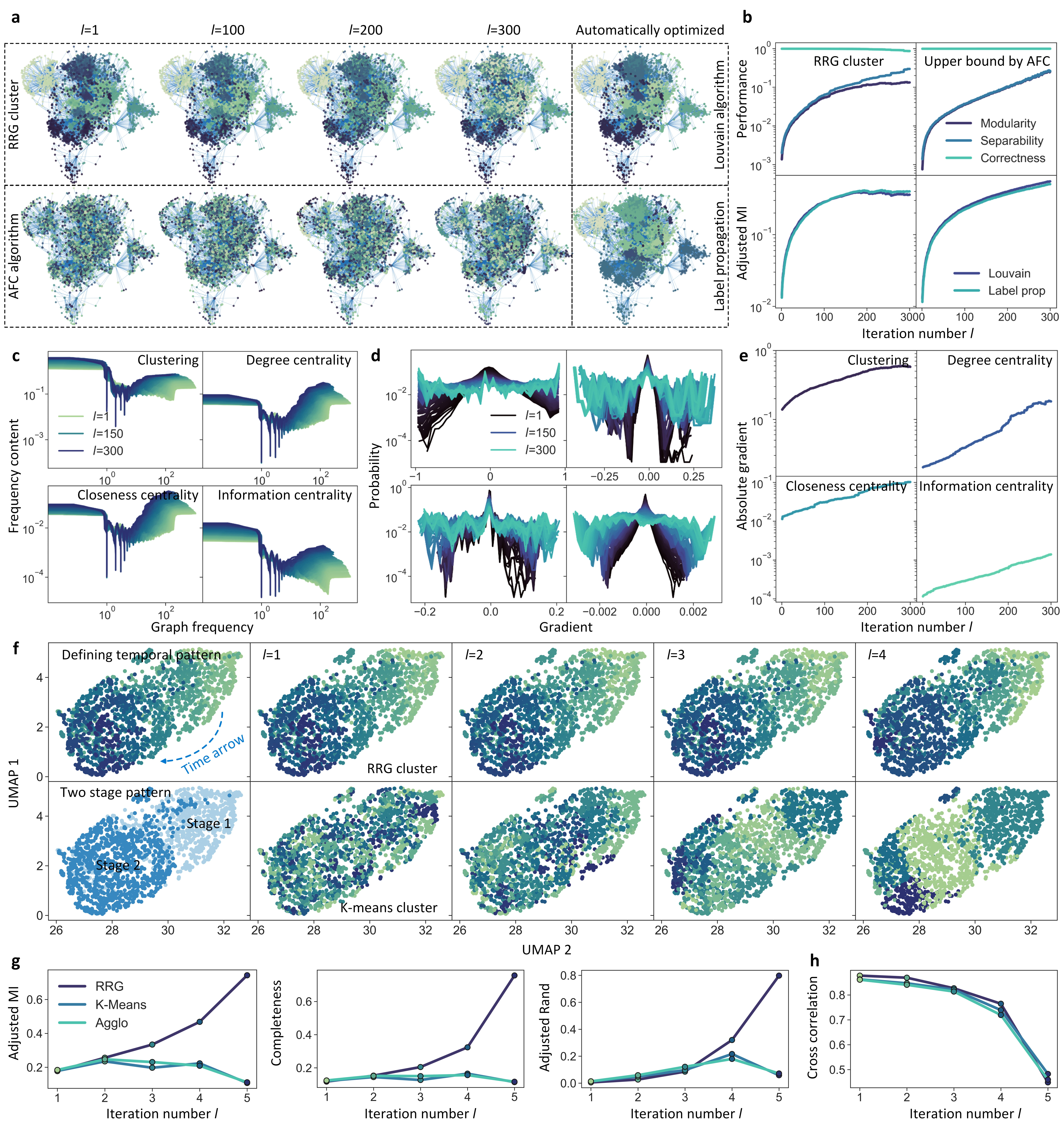}
\caption{\label{G3} The intrinsic structures and dynamics revealed by renormalization. \textbf{a}, The RRG clusters identified in each iteration of the renormalization flow are compared with the communities discovered by the asynchronous fluid community (AFC), the Louvain and the Label propagation algorithms. Different clusters or communities are distinguished according to their colors. \textbf{b}, The modularity, separability, and correctness are compared between RRG clusters and ACF communities. Meanwhile, the adjusted mutual information (MI) between RRG clusters (or ACF communities) and Louvain communities (or label propagation communities) is measured in each iteration of renormalization. \textbf{c}, The frequency spectra of four kinds of graph signals defined by unit attributes during renormalization. \textbf{d-e}, The gradients and mean absolute gradients of these graph signals. \textbf{f}, The first column shows the temporal evolution process of the commodity system and its two-stage patterns in an UMAP embedding space. The last four columns show RRG and K-means clusters derived during renormalization. \textbf{g}, Different consistency degrees measured between RRG clusters (or K-means and agglomerative clusters) and the two-stage pattern of the commodity system. \textbf{g}, The mean cross-correlation values of three kinds of clusters are measured during renormalization.} 
\end{figure*}

For neuroscience, we use the largest high resolution central brain connectome of the fruit fly, \emph{Drosophila melanogaster} (see Fig. \ref{G1}h) \cite{xu2020connectome}. This data also serves as a platform to demonstrate the applicability of the RRG to directed networks. We select neurons and synapses when the associated cell bodies are positioned precisely. The filtered data constructs a network of $23008$ neurons and $635761$ pairs of synaptic adjacent relations. As shown in Figs. \ref{G1}i-j, the topology of Drosophila brain network is nearly scale-invariant as its degree (i.e., the sum of input- and output- degrees) distribution does not sharply change during renormalization. 

For medical science, we find that the RRG helps to distinguish between cancer phenotypes (e.g., B and T-cell acute leukemia, acute myeloid leukemia, and multiple myeloma) and healthy bone marrow according to the renormalization flows of their gene co-expression networks \cite{nakamura2023network}. In Figs. \ref{G1}k-l, all gene co-expression networks of normal bone marrow consist of single connected components with rich connections, whose degree distributions change across different iterations of the RRG. As comparisons, their counterparts under cancer conditions feature more sparse connectivity structures and maintain more robust during renormalization. Consequently, the closeness to scale-invariance property of a gene co-expression network under the RRG transformation may serve as an auxiliary criterion to detect cancer phenotypes.


\section{Dynamics renormalization}
To validate the applicability of the RRG in dynamics renormalization, we apply it to a two-dimensional contact process \cite{marro2005nonequilibrium}, which belongs to the directed
percolation universality class \cite{saberi2015recent} (see Methods for data preparation). As the control parameter of contact process, the spreading rate, approaches to $1.649$, the contact process exhibits an absorbing phase transition \cite{marro2005nonequilibrium}. As the control parameter continues to increase, the contact process arrives in its super-critical phase. As shown in Fig. \ref{G2}a, a RRG with the same parameter settings as Fig. \ref{G1} successfully reveals and preserves the non-trivial correlation patterns (represented by correlation matrices) among units under both critical and super-critical conditions. In Fig. \ref{G2}b, the RRG distinguishes between these two conditions by showing a non-Gaussian fix point (i.e., with a high skewness) of the critical state and a Gaussian fix point (i.e., with a standard bell shape) of the super-critical state, which are consistent with the findings of Ref. \cite{nicoletti2020scaling}. In Figs. \ref{G2}c-g, different scaling features of the RRG are estimated from the data with high accuracy. Compared with the super-critical state, the critical state has a value of $\alpha$ closer to $1.5$, suggesting that the critical state is nearer to the midpoint between perfect independence and strong coupling. Meanwhile, for a contact process, a scaling relation $\eta=d-2+\beta/\nu_{\bot}$ exists among these exponents, where $\eta=2-d\mu$, notion $d=2$ measures the dimension, and $\nu_{\bot}\simeq 0.733$ \cite{marro2005nonequilibrium,dickman2005quasi}. As shown by the results of the critical state, the RRG derives $\eta\simeq 0.88$ and $d-2+\beta/\nu_{\bot}\simeq 0.9$, which is consistent with the scaling relation with reasonable errors. The RRG also distinguishes the super-critical case from the critical one by showing that $\eta=d-2+\beta/\nu_{\bot}$ fails with large deviations. Moreover, the critical state principally exhibits the universal collapse in Fig. \ref{G2}f, while the super-critical state does not. A dynamic scaling phenomenon of the critical state can be seen in Fig. \ref{G2}g. The super-critical case has a much smaller exponent $\theta$ (i.e., closer to zero), suggesting a nearly constant correlation time across different macro-unit sizes (therefore, there is no significant dynamic scaling). These results all validate the robust ability of the RRG to classify critical and non-critical dynamics.

As an application on complicated data, we implement the RRG on the whole-brain dynamics of zebrafish larva \cite{van2023neural} (all settings of the RRG keep the same as those in contact process analysis). Eight brains are analyzed, where the activities of $\simeq 40000$ neurons are recorded in each brain (see Methods). As shown in Fig. \ref{G2}h, the renormalization flows of all brains converge to certain non-Gaussian fixed points, rejecting the possibility that the whole-brain dynamics proceeds with weakly correlated or independent neurons. The properties of non-trivially correlated neural activities are reflected by scaling features, which are estimated from the data with small errors. Exponents $\alpha$ and $\beta$ generally suggest the non-negligible correlations among neurons (Figs. \ref{G2}i-j). These correlations are self-similar since a power-law scaling phenomenon depending on mean fractional rank, $r/\langle k\rangle$, exists in every covariance matrix eigenvalue spectrum as a consequence of scale-invariant propagator (Fig. \ref{G2}k). In Fig. \ref{G2}l, we observe the dynamic scaling of temporal correlations, whose exponents are numerically similar to the results of a phenomenological RG implemented on mouse brain dynamics. We see systematic differences between the scaling behaviours of the two-dimensional contact process and the whole-brain dynamics, which may arise from the widespread long-range interactions among neurons (e.g., the small-world connectivity). The RRG is suitable for analyzing the dynamics with long-range interactions because its coarse graining for dynamics is independent of initial spatial structures (certainly, spatial constraints can be easily added into $Y^{\left(l\right)}$ if necessary).

\section{Intrinsic structures and dynamics discovered by the RRG}

Why is the RRG effective in structure and dynamics renormalization? In Fig. \ref{G3}, we show that the effectiveness may arise from the ability of the RRG to unveil the intrinsic structures and dynamics of target systems.

For structure renormalization, each iteration of the RRG is demonstrated to discover an ideal partition available under the corresponding condition. In Fig. \ref{G3}a, we use the Facebook data \cite{FacebookDataBase} in Figs. \ref{G1}f-g as an instance. We implement an RRG with the signed Cauchy projection and a hashed representation, $Z^{\left(l\right)}$, whose dimension is $100$ on the data for $300$ iterations. In the $l$-th iteration, we count the number of macro-units, $N^{\left(l\right)}$, and apply the asynchronous fluid community algorithm \cite{pares2018fluid} to identify $N^{\left(l\right)}$ communities on the network. Meanwhile, we generate another kind of communities directly using the RRG. For each macro-unit in $X^{\left(l\right)}$, we define the set of all initial units in $X^{\left(1\right)}$ aggregated into this macro-unit as an RRG cluster. Moreover, we apply the Louvain algorithm \cite{blondel2008fast} and the label propagation algorithm \cite{gregory2010finding} to automatically detect optimal partitions of the network, where the optimized community quantities are not necessarily the same as the RRG. In Fig. \ref{G3}a, compared with asynchronous fluid communities, the partition formed by RRG clusters is more qualitatively similar to the ones formed by Louvain and label propagation communities in each iteration. Even though the RRG has never been optimized using common  partition quality metrics (e.g., modularity, separability, and correctness, whose definitions are presented in Methods), it robustly achieves competitive performance as the asynchronous fluid community algorithm at each time of network partition (Fig. \ref{G3}b). As the RRG iterates, the adjusted mutual information (see Methods) between the partition created by the RRG and those implied by Louvain and label propagation communities progressively increases (Fig. \ref{G3}a), suggesting that the RRG essentially drives the system to its intrinsic structure with an optimal partition that is similar with but in-equivalent to those derived by community detection algorithms.

To quantitatively understand the effects of the RRG on structures, we measure four representative attributes (e.g., clustering coefficient \cite{saramaki2007generalizations} and three kinds of centrality metrics \cite{borgatti2005centrality,freeman2002centrality,brandes2005centrality}, see Methods) on every unit in $X^{\left(l\right)}$. Each attribute defines a graph signal distributed on the network, whose frequency representation is derived by the graph Fourier transform \cite{ortega2018graph,dong2020graph} (Fig. \ref{G3}c and Methods). As the RRG iterates, the frequency spectra of all graph signals have increasingly more contents (i.e., energies) concentrated in high frequencies (i.e., graph signals become un-smooth), suggesting the enlarged differences between macro-units. These growing differences can also be validated by the probability distributions of graph signal gradients \cite{ortega2018graph,dong2020graph} (Fig. \ref{G3}d), and the mean absolute gradient (Fig. \ref{G3}e), where the probability masses of non-zero gradients increase during renormalization and the mean absolute gradient persistently grows. In sum, the RRG reduces the redundant information shared by strongly correlated units in every iteration. Each generated macro-unit becomes increasingly unique, implying a structure with lower redundancy and higher modularity.

For dynamics renormalization, we analyze the RRG following a similar paradigm. We use the long-term consumption dynamics data of about 6000 kinds of commodities collected in one of our earlier works \cite{ConsumptionDataBase}. The data is extracted from approximately 2.2 million purchase orders made from January 2018 to December 2022 (see Methods). The UMAP embedding \cite{mcinnes2018umap} of this data reveals a clear temporal evolution process of the commodity system (Fig. \ref{G3}f), which can be further subdivided into two stages separated by the outbreak of COVID-19 (i.e., before and after the outbreak) \cite{ConsumptionDataBase}. We use an RRG with the signed Cauchy projection (the dimension of $Z^{\left(l\right)}$ is $1000$) to renormalize the system on the time dimension (i.e., each time step serves as a unit) and define RRG clusters according to unit aggregation. The derived $N^{\left(l\right)}$ RRG clusters in the $l$-th iteration define a temporal evolution pattern. Meanwhile, a K-means clustering \cite{hartigan1979algorithm,ashabi2020systematic} and an agglomerative clustering algorithms \cite{mullner2011modern} are used to partition the time domain into $N^{\left(l\right)}$ clusters. As shown in Fig. \ref{G3}f, the RRG is better at recovering the actual temporal evolution process than K-means and agglomerative clustering algorithms, which can be quantitatively validated by the consistency \cite{vinh2009information,rosenberg2007v,steinley2004properties} between the discovered temporal evolution and the actual two-stage pattern (Fig. \ref{G3}g). Finally, after measuring the mean cross correlation (i.e., the correlations among units in different clusters), the RRG is suggested to make macro-units increasingly different from each other, which is consistent with our findings in Figs. \ref{G3}c-e.

\section{Acceleration and memory saving}

 \begin{figure*}[!t]
\includegraphics[width=1\columnwidth]{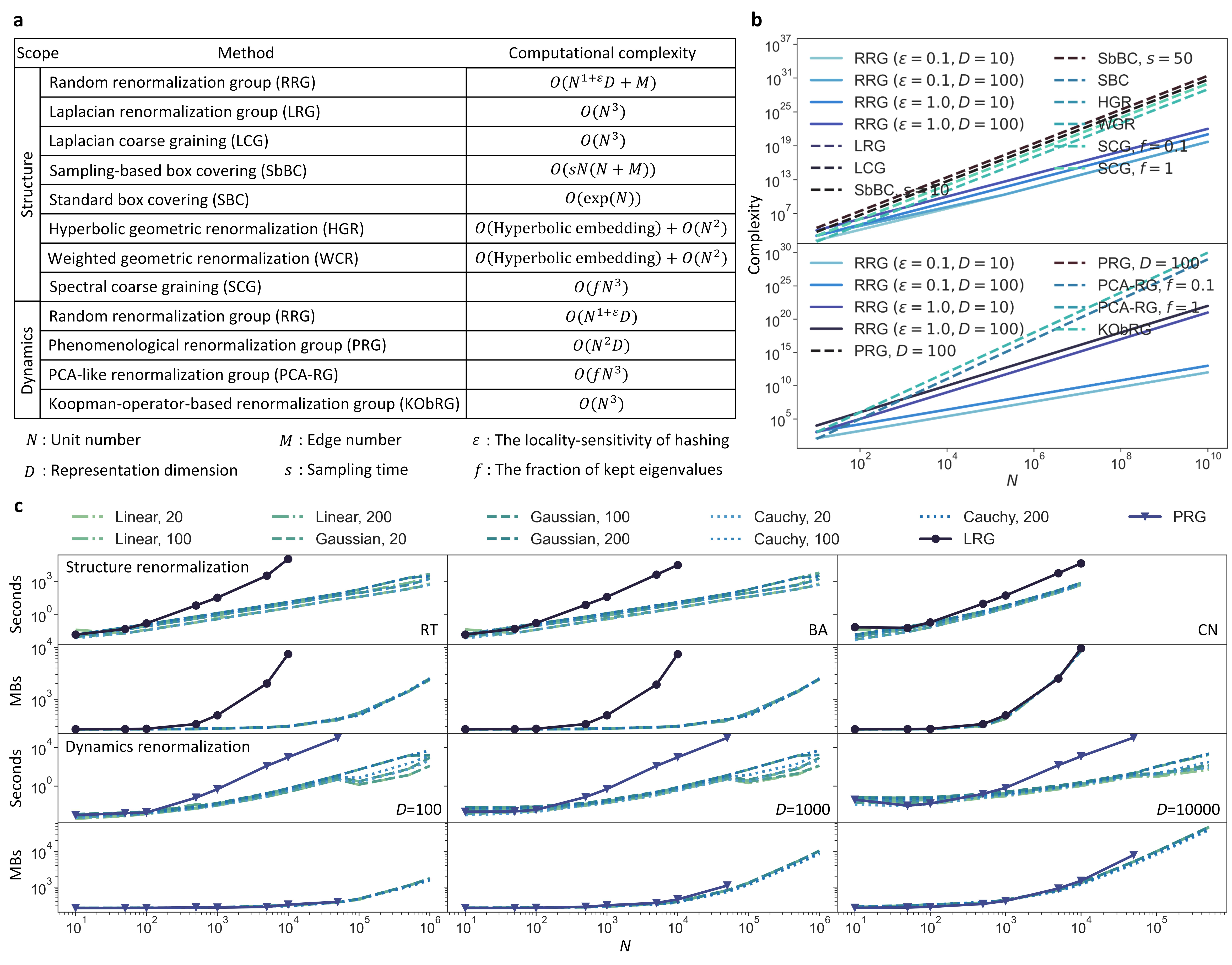}
\caption{\label{G4} The computational efficiency of the RRG. \textbf{a}, The computational complexities of the RRG and other existing RG frameworks in structure and dynamics renormalization. The considered RG frameworks include the Laplacian RG \cite{villegas2023laplacian}, the Laplacian coarse graining \cite{loures2023laplacian}, the sampling-based \cite{wei2019sampling} and the standard box covering \cite{song2005self,song2007calculate}, the hyperbolic geometric renormalization \cite{garcia2018multiscale} and its weighted network variant \cite{zheng2023geometric}, the spectral coarse graining \cite{gfeller2007spectral}, the phenomenological RG \cite{nicoletti2020scaling}, the PCA-like RG \cite{bradde2017pca}, and the Koopman-operator-based RG \cite{redman2020renormalization}. \textbf{b}, The growing trends of different computational complexities in \textbf{a} are illustrated in the worst case. For structure renormalization, the worst case denotes an extreme condition where system structures are complete networks (i.e., units share all-to-all connections). Note that the computational complexity of the hyperbolic embedding is measured following Refs. \cite{papadopoulos2015network,blasius2018efficient}. \textbf{c}, The one-step-computation time and memory costs of the RRG and its representative competitors are measured under different conditions (e.g., notion ``Linear, 20" denotes the signed random hyperplane projection and a binary representation, $Z^{\left(l\right)}_{i}$, whose dimension is 20). For structure renormalization, the random tree (RT), the Barab{\'a}si-Albert network (BA) \cite{albert2002statistical}, and the complete network (CN) are considered. Note that some data points of classic RG frameworks are missing due to memory errors (i.e., out of memory).} 
\end{figure*}

    As previously mentioned, the RRG is rooted in random projections, hashing techniques, and kernel representations. These designs ensure the processing capacity of diverse complicated correlations and significantly reduce computational costs. In terms of computational complexity, we compare the RRG with other optimization-free RG frameworks in structure and dynamics renormalization (see Fig. \ref{G4}a for complexities and see Fig. \ref{G4}b for the growth trends of complexities in the worst case). As a function of unit number $N$, edge number $M$, representation dimension (i.e., the dimension of $Y^{\left(l\right)}$ in step (1)) $D$, and hashing locality sensitivity $\varepsilon\in\left(0,1\right)$ (determined by the dimension of $Z^{\left(l\right)}$ in step (3) and can be understood as the precision of approximating a kernel space by random projections), the complexity of the RRG is smaller than all considered competitors (see Methods). To validate this advantage in real cases, we compare between the RRG and two representative RGs according to time and memory costs (see Methods). Different kernel and parameter designs of the RRG are considered, which are all proven as efficient in time and memory saving. These results demonstrate the RRG as a computationally friendly framework for ultra-large system processing in personal computer environments.

\section{Discussion}

It is the computation that bridges between theories and experiments in physics \cite{hendrickson2009computational}. Even back in the Kepler age (i.e., the 17th century), computation (manually or using slide rules) on recorded data has become the key way to reject hypotheses and discover new theories. Since the 20th century, the flourishing developments of electronic techniques and computers have given new missions to computation, i.e., dealing with the ultra-large scale data generated by known or unknown mechanisms to reveal underlying mechanics with acceptable computational costs. In these missions, the statistical modelling ability of complex relations becomes the key towards latent mechanism identification. Time- and memory-efficiency become the bottlenecks of discovering new physics in real data. These emerging trends have reformed diverse physics fields, especially condensed matter physics \cite{ohno2018computational,bedolla2020machine,schmidt2019recent,schleder2019dft}, fluid mechanics \cite{ferziger2019computational,brunton2020machine,anderson2020computational}, quantum mechanics \cite{mcardle2020quantum,bauer2020quantum,cao2019quantum}, and astronomy \cite{borne2009astroinformatics,almgren2013nyx,anninos2001computational}.

In this work, we have brought this reformation to the renormalization group, a theory that analyzes criticality, symmetry, and various important physics concepts \cite{pelissetto2002critical,jona2001renormalization}. To meet the different demands of studying complex systems, we equip the RRG with a general ability to realize both structure and dynamics renormalization under a unified framework. This framework consists of multiple carefully designed computational mathematics components and serves as an end-to-end pipeline without any requirement on \emph{a priori} knowledge (Fig. \ref{G0}). For system structures, it integrates out short-range correlations in local region topology properties to select long-range correlations and identifies potential scale-invariance (Fig. \ref{G1}). For system dynamics, it enables us to verify diverse scaling behaviours to explore the existence of criticality (Fig. \ref{G2}). The effectiveness of these features is ensured by the robust capacity of the RRG to revel intrinsic system structures and dynamics (Fig. \ref{G3}). Although kernel designs \cite{hofmann2008kernel,shawe2004kernel} are included in the RRG to guarantee the modelling capacity of different unit correlations, the modelled correlation patterns in kernel spaces can be efficiently approximated using random projections \cite{vempala2005random,li2022signrff,li2013sign,rahimi2007random}, hashing functions \cite{broder1997resemblance,broder1998min}, and approximate nearest neighbor search \cite{liu2004investigation,li2019approximate}. Therefore, the RRG can realize significant reductions in time and memory cost compared with existing works \cite{villegas2023laplacian,loures2023laplacian,wei2019sampling,song2005self,song2007calculate,garcia2018multiscale,zheng2023geometric,gfeller2007spectral,nicoletti2020scaling,bradde2017pca,redman2020renormalization} (Fig. \ref{G4}).

Our motivation for proposing the RRG is to demonstrate the possibility to overcome the challenges faced by classic RG frameworks (i.e., the trade-offs between high computational complexity and sufficient modelling capacities of complex systems). This demonstration serves as a starting point to explore more possibilities in future studies. As we have discussed, optimization-free \cite{villegas2023laplacian,loures2023laplacian,wei2019sampling,song2005self,song2007calculate,garcia2018multiscale,zheng2023geometric,gfeller2007spectral,nicoletti2020scaling,bradde2017pca,redman2020renormalization} and optimization-dependent RGs \cite{ron2017surprising,troster2015fourier,wu2019determination,ron2002inverse,wu2017variational,wu2020continuous,bachtis2021adding,di2022deep,koch2018mutual,li2018neural,chung2021neural,hu2022rg} have their own advantages and disadvantages in solving physics questions. While we have shown a scheme towards ideal optimization-free RG designs in the present work, there is no reason to ignore the possibility of combining the advantages of both kinds of RGs. As a suggestion, we advise paying attention to pre-trained large-scale sequence modelling learners (e.g., large language models) in deep learning \cite{gu2023mamba,gu2021efficiently,smith2022simplified,touvron2023llama}, which have been proven as optimal in information compression \cite{deletang2023language} and can be theoretically related to RGs in terms of information bottleneck theory \cite{kline2022gaussian,gordon2021relevance,lenggenhager2020optimal}. Notably, many kernels, random projections, and hashing techniques can be naturally included in these learners. Therefore, future explorations of information-compressor-based RGs may also benefit from our work in technical aspects.

 \bibliography{Main}
 
\appendix
\section*{Methods}\label{ASec-1}

\subsection{The random renormalization group framework}
In this section, we elaborate the theoretical framework of the RRG. Let us consider a system $X=\left(X_{1},\ldots,X_{N}\right)$ of size $N$, it is expected as a network of $N$ units for structure renormalization or a set of $N$ sequences for dynamics renormalization. The iterations of the RRG begin with setting $X^{\left(1\right)}=X$, after which each $l$-th iteration uses $X^{\left(l\right)}$ as the input to derive $X^{\left(l+1\right)}$. Every iteration of the RRG follows the same pipeline to renormalize the system. Below, we introduce this pipeline in details: 

\paragraph*{\textbf{Step (1)}} Given $X^{\left(l\right)}$, we first define its feature representation, $Y^{\left(l\right)}$, such that the concerned properties of every unit $X^{\left(l\right)}_{i}$ are described by a feature vector $Y^{\left(l\right)}_{i}$.  
    
For structure renormalization, each vector $Y^{\left(l\right)}_{i}$ is defined as
\begin{align}
Y^{\left(l\right)}_{i}=\operatorname{MinHash}\left(W^{\left(l\right)}_{i},\;D\right).\label{EQ1}
\end{align}
In Eq. (\ref{EQ1}), notion $W^{\left(l\right)}_{i}$ denotes the set consisting of $X^{\left(l\right)}_{i}$ and all its adjacent units. Function $\operatorname{MinHash}\left(\cdot,\;D\right)$ is the MinHash mapping \cite{broder1997resemblance,broder1998min} that transforms $W^{\left(l\right)}_{i}$ to a vector of $D$-dimension. The MinHash ensures that the Jaccard distance between $W^{\left(l\right)}_{i}$ and $W^{\left(l\right)}_{j}$ (i.e., the difference between the local adjacent relations concerning units $X^{\left(l\right)}_{i}$ and $X^{\left(l\right)}_{j}$) can be approximated by the normalized XOR distance between $Y^{\left(l\right)}_{i}$ and $Y^{\left(l\right)}_{j}$
\begin{align}
J\left( W^{\left(l\right)}_{i},W^{\left(l\right)}_{j}\right)=\lim_{D\rightarrow\infty}\frac{1}{D}\sum_{k=1}^{D} Y^{\left(l\right)}_{i}\left(k\right) \oplus Y^{\left(l\right)}_{j}\left(k\right),\label{EQ2}
\end{align}
where $\oplus$ denotes the exclusive disjunction operator and $Y^{\left(l\right)}_{i}\left(k\right)$ denotes the $k$-th component of $Y^{\left(l\right)}_{i}$. 

For dynamics renormalization, there is no need for extra processing since the dynamics sequence associated with unit $X^{\left(l\right)}_{i}$ already serves as a feature vector $Y^{\left(l\right)}_{i}$. 

To enhance numerical stability, we normalize each $Y^{\left(l\right)}_{i}$ before subsequent processing. For dynamics renormalization, the normalization of $Y^{\left(l\right)}_{i}$ is designed to control the numerical effects of extreme values (e.g., orders of magnitude) in dynamics sequences. Specifically, each $Y^{\left(l\right)}_{i}$ is normalized to have a zero mean for the signed random hyperplane projection. For the signed random Fourier feature, each $Y^{\left(l\right)}_{i}$ is normalized to have a zero mean and unit variance. For the signed Cauchy projection, we normalize $Y^{\left(l\right)}_{i}$ such that $\sum_{k=1}^{D} Y^{\left(l\right)}_{i}\left(k\right)=1$. For structure renormalization, there are fewer extreme value issues faced by the normalized XOR distance after the MinHash transformation. Meanwhile, the normalized XOR distance may be significantly disturbed by normalization. Therefore, we define the normalization as an identical mapping and use $Y^{\left(l\right)}_{i}$ as the normalized result directly. Note that the existence of normalization does not change the nature of kernel spaces. The normalization procedures for dynamics renormalization are defined following Refs. \cite{vempala2005random,li2022signrff,li2013sign} while the identical mapping for structure renormalization makes these kernel spaces be defined on $L_{1}$ or $L_{2}$ norms as described in Ref. \cite{rahimi2007random}. 

\paragraph*{\textbf{Step (2)}} After deriving the feature representation, we measure the correlation between any pair of units, $X^{\left(l\right)}_{i}$ and $X^{\left(l\right)}_{j}$, based on $Y^{\left(l\right)}_{i}$ and $Y^{\left(l\right)}_{j}$. The correlation can be defined in a linear or non-linear manner and three possible choices are offered in the RRG. 

The first choice is related to the cosine similarity in a linear kernel space, which behaves like the linear correlation. We introduce the signed random hyperplane projection \cite{vempala2005random}
\begin{align}
Z^{\left(l\right)}_{i}=\frac{1}{2}\left(1+\operatorname{sign}\left(Y^{\left(l\right)}_{i}\omega\right)\right),\label{EQ3}
\end{align}
where $\omega$ is a $\left(d\times h\right)$-dimensional random matrix such that each matrix element $\omega_{pq}$ is a standard Gaussian variable. For structure renormalization, dimension $d=D$ is set to enable the matrix multiplication. For dynamics renormalization, dimension $d$ equals to the length of dynamics sequence. Parameter $h$ determines the dimension of $Z^{\left(l\right)}_{i}$ and is adjustable in the RRG. The projected $Z^{\left(l\right)}_{i}\in\{0,1\}^{h}$ is a binary vector. The Hamming distance, $H\left(i,j\right)$, reflects the cosine similarity between $X^{\left(l\right)}_{i}$ and $X^{\left(l\right)}_{j}$ because the probability for $H\left(i,j\right)$ to equal $s\leq h$ is given as 
\begin{align}
&P\left(H\left(i,j\right)=s\right)\notag\\=&\binom{h}{s}\left(\frac{\arccos\left(\rho\right)}{\pi}\right)^{s}\left(1-\frac{\arccos\left(\rho\right)}{\pi}\right)^{h-s},\label{EQ4}
\end{align}
where $\rho=\left(Y^{\left(l\right)}_{i}\right)^{\top}Y^{\left(l\right)}_{j}$ is the cosine similarity. As $\rho$ enlarges, the probability defined in Eq. (\ref{EQ4}) decreases for a large $s$ and increases for the $s$ approaching to zero.

The second choice is the signed random Fourier feature with a Gaussian kernel, which is derived by 
\begin{align}
Z^{\left(l\right)}_{i}=\frac{1}{2}\left(1+\operatorname{sign}\left(\cos\left(Y^{\left(l\right)}_{i}\omega+\psi\right)\right)\right)\label{EQ5}
\end{align}
using a random vector whose components are uniformly distributed in $\left[0,2\pi\right]$ \cite{li2022signrff}. Based on this setting, the Hamming distance, $H\left(i,j\right)$, is related to the similarity between $X^{\left(l\right)}_{i}$ and $X^{\left(l\right)}_{j}$ in the Gaussian kernel space
\begin{align}
P\left(H\left(i,j\right)=s\right)=\binom{h}{s}\left(1-P_{\text{col}}\right)^{s}\left(P_{\text{col}}\right)^{h-s},\label{EQ6}
\end{align}
where notion
\begin{align}
    P_{\text{col}}=\int_{0}^{1}\int_{0}^{1}f_{\rho}\left( R^{\left(l\right)}_{i},R^{\left(l\right)}_{j}\right)\mathsf{d}R^{\left(l\right)}_{i}\mathsf{d}R^{\left(l\right)}_{j}\label{EQ7}
\end{align}
defines the collision probability using the random Fourier feature $R^{\left(l\right)}_{i}=\cos\left(Y^{\left(l\right)}_{i}\omega+\psi\right)$. The joint distribution of two random Fourier features in Eq. (\ref{EQ7}) is 
\begin{align}
    f_{\rho}\left( R^{\left(l\right)}_{i},R^{\left(l\right)}_{j}\right)=\frac{\sum\limits_{r=-\infty}^{\infty}g_{\rho}\left(A^{\left(l\right)}_{ij}+2r\pi\right) +g_{\rho}\left(B^{\left(l\right)}_{ij}+2r\pi\right)}{\pi\sqrt{1-\left(R^{\left(l\right)}_{i}\right)^{2}}\sqrt{1-\left(R^{\left(l\right)}_{j}\right)^{2}}},\label{EQ8}
\end{align}
where we mark $A^{\left(l\right)}_{i}=\arccos\left(R^{\left(l\right)}_{i}\right)-\arccos\left(R^{\left(l\right)}_{j}\right)$ and $B^{\left(l\right)}_{i}=\arccos\left(R^{\left(l\right)}_{i}\right)+\arccos\left(R^{\left(l\right)}_{j}\right)$. Notion $g_{\rho}\left(\cdot\right)$ denotes the probability density function of a Guassian variable whose mean is zero and variance is $2-2\rho$. As suggested in Ref. \cite{li2022signrff}, Eqs. (\ref{EQ7}-\ref{EQ8}) lead to that $\mathbb{E}\left[\operatorname{sign}\left(R^{\left(l\right)}_{i}\left(k\right)\right)\operatorname{sign}\left(R^{\left(l\right)}_{j}\left(k\right)\right)\right]$ monotonously increases with $\rho$ for each $k$-th component ($k\in\{1,\ldots,h\}$). Because probability distributions are invariant under the linear shift in Eq. (\ref{EQ5}), we know that $\mathbb{E}\left[Z^{\left(l\right)}_{i}\left(k\right)Z^{\left(l\right)}_{j}\left(k\right)\right]$ and $P_{\text{col}}$ are monotonously increasing functions of $\rho$ as well. Therefore, the probability $P\left(H\left(i,j\right)=s\right)$ associated with a small $s$ enlarges with $\rho$.

The third choice is the similarity between $X^{\left(l\right)}_{i}$ and $X^{\left(l\right)}_{j}$ in a Cauchy kernel space \cite{li2013sign}. Formally, this Cauchy similarity is defined as $\rho_{\chi^{2}}=\sum_{k=1}^{h}\frac{2 Y^{\left(l\right)}_{i}\left(k\right)Y^{\left(l\right)}_{j}\left(k\right)}{Y^{\left(l\right)}_{i}\left(k\right)+Y^{\left(l\right)}_{j}\left(k\right)}$, to represent which, we define
\begin{align}
Z^{\left(l\right)}_{i}=\frac{1}{2}\left(1+\operatorname{sign}\left(\cos\left(Y^{\left(l\right)}_{i}\phi+\psi\right)\right)\right),\label{EQ9}
\end{align}
where $\phi$ is a $\left(d\times h\right)$-dimensional random matrix such that each element $\phi_{pq}$ is a Cauchy variable with zero mean and unit variance. The Hamming distance, $H\left(i,j\right)$, can reflect the Cauchy similarity because
\begin{align}
&P\left(H\left(i,j\right)=s\right)\notag\\\simeq&\binom{h}{s}\left(\frac{\arccos\left(\rho_{\chi^{2}}\right)}{\pi}\right)^{s}\left(1-\frac{\arccos\left(\rho_{\chi^{2}}\right)}{\pi}\right)^{h-s}.\label{EQ10}
\end{align}
Given a larger Cauchy similarity, the probability for $H\left(i,j\right)$ to equal to a large $s$ becomes smaller.

\paragraph*{\textbf{Step (3)}} Because the Hamming distance between $Z^{\left(l\right)}_{i}$ and $Z^{\left(l\right)}_{j}$ can approximate different kinds of kernel correlations between $X^{\left(l\right)}_{i}$ and $X^{\left(l\right)}_{j}$, we can primarily focus on the Hamming space of $Z^{\left(l\right)}$ rather than those computationally expensive kernel spaces. 

Since our goal is to select short-range correlations in the system (i.e., find highly similar units), there is no need to compute the Hamming distance between any pair of $Z^{\left(l\right)}_{i}$ and $Z^{\left(l\right)}_{j}$. Instead, we can implement the approximate nearest neighbor search \cite{liu2004investigation,li2019approximate} on $Z^{\left(l\right)}$, which naturally fits in with the Hamming distance among binary variables. In the RRG, we use the Faiss \cite{douze2024faiss}, an efficient C++ library for similarity search in large-scale database, to construct $U^{\left(l\right)}$, the space containing all searched nearest neighbor relations among units. Specifically, we realize an exhaustive search to ensure a high recall when the umber of units is no more than $5\times 10^{4}$. If the system size is greater than $5\times 10^{4}$ and no more than $5\times 10^{5}$, we implement a search with bucket quantization for acceleration. When the system contains more than $5\times 10^{5}$ units, we design an approximate search with bucket quantization and only use the first $h^{\prime}$ dimensions of each $Z^{\left(l\right)}_{i}$, where $h^{\prime}=\min\{\max\{0.01h,32\},16\}$ denotes an empirical dimension cut-off.

\paragraph*{\textbf{Step (4)}} Then, we use the derived space $U^{\left(l\right)}$ to define a nearest neighbor network. We begin with a null network, $G^{\left(l\right)}$, containing all units and progressively add new edges into it. For dynamics renormalization, we add an edge between units $X^{\left(l\right)}_{i}$ and $X^{\left(l\right)}_{j}$ in $G^{\left(l\right)}$ if they are nearest neighbors in $U^{\left(l\right)}$. For structure renormalization, we connect between $X^{\left(l\right)}_{i}$ and $X^{\left(l\right)}_{j}$ in $G^{\left(l\right)}$ only when these two units are adjacent both in $X^{\left(l\right)}$ and $U^{\left(l\right)}$. After dealing with every pair of units, we form multiple connected clusters in $G^{\left(l\right)}$, where each connected cluster, $C^{\left(l\right)}_{k}$, contains a set of units sharing strong correlations.

\paragraph*{\textbf{Step (5)}} Finally, we renormalize the system by coarse graining the units in each connected cluster $C^{\left(l\right)}_{k}$ into a macro-unit $X^{\left(l+1\right)}_{k}$. For dynamics renormalization, every macro-unit $X^{\left(l+1\right)}_{k}$ is defined with a summed feature vector $Y^{\left(l+1\right)}_{k}=\sum_{i\in I^{\left(l\right)}_{k}} Y^{\left(l\right)}_{i}$, where $I^{\left(l\right)}_{k}$ is the index set of all units contained in cluster $C^{\left(l\right)}_{k}$. For structure renormalization, two macro-units, $X^{\left(l+1\right)}_{i}$ and $X^{\left(l+1\right)}_{j}$, are connected in $X^{\left(l+1\right)}$ if the units aggregated into $X^{\left(l+1\right)}_{i}$ share at least one edge with the units aggregated into $X^{\left(l+1\right)}_{j}$. 

By repeating steps (1-5) for $T$ iterations, we can generate a sequence of the system on different scales, $\left[X^{\left(1\right)},\ldots,X^{\left(T\right)}\right]$, which is referred to as the renormalization flow. 

Please note that the feature representation of the system is always $Y^{\left(l\right)}$ rather than $Z^{\left(l\right)}$. The hashed result $Z^{\left(l\right)}$ is only to select short-range correlations to guide renormalization. 

\subsection{Scaling behaviours and macroscopic observables}
In this section, we introduce the macroscopic observables and scaling features used for analyzing the RRG. 

\paragraph*{\textbf{Macroscopic observables}} For structure renormalization, the macroscopic observable is defined as the degree distribution of units 
\begin{align}
P^{\left(l\right)}\left(\operatorname{deg}=n\right)\simeq\frac{1}{N^{\left(l\right)}}\sum_{X^{\left(l\right)}_{i}}\delta\left(\operatorname{deg}\left(X^{\left(l\right)}_{i}\right),n\right),\label{EQ11}
\end{align}
where $N^{\left(l\right)}$ counts the number of units in $X^{\left(l\right)}$, notion $\operatorname{deg}\left(\cdot\right)$ denotes the degree, and $\delta\left(\cdot,\cdot\right)$ is the Kronecker delta function. We can use the Kolmogorov–Smirnov statistic \cite{simard2011computing,berger2014kolmogorov} between $P^{\left(l\right)}\left(\cdot\right)$ and $P^{\left(1\right)}\left(\cdot\right)$ to measure how the normalized system departs from its original state ($l\geq 2$)
\begin{align}
\Delta\left(l,1\right)=\sup_{n}\vert P^{\left(l\right)}\left(\operatorname{deg}=n\right)-P^{\left(1\right)}\left(\operatorname{deg}=n\right)\vert,\label{EQ12}
\end{align}
whose statistical significance, $p^{\left(l\right)}$, can be calculated by the two-sided Kolmogorov–Smirnov test \cite{simard2011computing,berger2014kolmogorov} given a null hypothesis that $P^{\left(l\right)}\left(\operatorname{deg}=n\right)=P^{\left(1\right)}\left(\operatorname{deg}=n\right)$ holds for any $n$. We can further average Eq. (\ref{EQ12}) across iterations to derive the mean evolution intensity of the system during renormalization
\begin{align}
\operatorname{KS}\left(T\right)=\frac{1}{T-1}\sum_{l=2}^{T}\Theta\left(0.01-p^{\left(l\right)}\right)\Delta\left(l,1\right),\label{EQ13}
\end{align}
where $\Theta\left(\cdot\right)$ denotes the unit step function (i.e., $\Theta\left(x\right)=1$ if $x\geq 0$ and $\Theta\left(x\right)=0$ otherwise) and $0.01$ denotes a hard threshold of statistical significance. When $p^{\left(l\right)}\leq 0.01$, it is safe to reject the null hypothesis and treat $P^{\left(l\right)}\left(\cdot\right)$ as different from $P^{\left(1\right)}\left(\cdot\right)$ significantly. Otherwise, the null hypothesis can not be rejected and $\Delta\left(l,1\right)$ may arise from numerical errors or noises. In practice, we can use $\operatorname{KS}\left(T\right)$ to determine whether the degree distribution is invariant under the RRG transformation (i.e., scale-invariance). In the case where the renormalization flow is short (e.g., when $T\leq 10$), we suggest $0.05$ as an empirical standard to determine scale-invariance. When the renormalization flow is long (e.g., when $T\geq 50$), we use $0.1$ or $0.15$ as a threshold of $\operatorname{KS}\left(T\right)$. We set a larger threshold for the long renormalization flow because $N^{\left(l\right)}$ in Eq. (\ref{EQ11}) may become sufficiently small as $l$ enlarges. In this case, the normalized frequency distribution on the right side of Eq. (\ref{EQ11}) approximate the actual probability distribution on the left side of Eq. (\ref{EQ11}) with large errors. An appropriate threshold should be robust enough to tolerate these noises and, therefore, becomes larger.

For dynamics renormalization, we denote the probability distribution of normalized dynamics as the macroscopic observable
\begin{align}
P^{\left(l\right)}\left(V=v\right)\simeq\frac{1}{N^{\left(l\right)}}\sum_{V^{\left(l\right)}_{i}}\delta\left(V^{\left(l\right)}_{i},v\right),\label{EQ14}
\end{align}
where $V^{\left(l\right)}_{i}$ denotes the normalized dynamics of unit $X^{\left(l\right)}_{i}$. We follow the idea of Ref. \cite{meshulam2019coarse} to define $V^{\left(l\right)}_{i}$ as
\begin{align}
V^{\left(l\right)}_{i}=\operatorname{std}\left(Q^{\left(l\right)}_{i}\right)^{-1}Q^{\left(l\right)}_{i},\label{EQ15}
\end{align}
where $\operatorname{std}\left(\cdot\right)$ denotes the standard deviation and $Q^{\left(l\right)}_{i}$ is the $i$-th row of matrix $Q^{\left(l\right)}$
\begin{align}
Q^{\left(l\right)}=\left(\sum_{o=1}^{O^{\left(l\right)}}L_{o}^{\top}L_{o}\right)\left(Y^{\left(l\right)}-\left[\mu^{\left(l\right)}_{1},\ldots, \mu^{\left(l\right)}_{N^{\left(l\right)}}\right]^{\top}\mathbf{1}\right).\label{EQ16}
\end{align}
In Eq. (\ref{EQ16}), notion $L_{o}$ denotes the eigenvector corresponding to the $o$-th largest eigenvalue of the covariance matrix formed among the feature vectors of all units in $X^{\left(l\right)}$ (i.e., the $\left(i,j\right)$-element of this matrix denotes the covariance between $Y^{\left(l\right)}_{i}$ and $Y^{\left(l\right)}_{j}$). Each term $\mu^{\left(l\right)}_{i}$ denotes the mean value of $Y^{\left(l\right)}_{i}$. Notion $\mathbf{1}$ denotes an all-one vector of an appropriate size.

The constructed matrix, $\sum_{o=1}^{O^{\left(l\right)}}L_{o}^{\top}L_{o}$, serves as a projector with a cut-off in the moment space when $O^{\left(l\right)}< N^{\left(l\right)}$. This cut-off excludes the contributions with low variances from the projector. After being projected in Eq. (\ref{EQ16}) and re-scaled in Eq. (\ref{EQ15}), the derived $V^{\left(l\right)}_{i}$ becomes the normalized dynamics of unit $X^{\left(l\right)}_{i}$ with a unit variance. In our work, we define $O^{\left(l\right)}=N^{\left(l\right)}/10$ for convenience. When the system exhibits critical dynamics, the probability distribution in Eq. (\ref{EQ14}) is expected to converge to a non-Gaussian fix point. When the dynamics is not critical (e.g., the correlations among units are too weak), the probability distribution of normalized dynamics may have a Gaussian fix point implied by the central limit theorem \cite{renyi2007probability} or lack a fix point.

\paragraph*{\textbf{Scaling behaviours}} For dynamics renormalization, the RRG also supports to analyze a series of scaling behaviours generalized from the phenomenological RG \cite{nicoletti2020scaling}.

First, we can study the mean variance of unit dynamics (i.e., the variance averaged across units)
\begin{align}
    \operatorname{Var}\left( \langle K^{\left(l\right)} \rangle\right)=\frac{1}{N^{\left(l\right)}}\sum_{i=1}^{N^{\left(l\right)}}\left[\nu^{\left(l\right)}_{i}-\left(\mu^{\left(l\right)}_{i}\right)^{2}\right],\label{EQ17}
\end{align}
where $\langle K^{\left(l\right)}\rangle$ measures the average size of units in $X^{\left(l\right)}$ and each $\nu^{\left(l\right)}_{i}$ denotes the mean of $\left(Y^{\left(l\right)}_{i}\right)^{2}$. We can calculate the sequences of $\operatorname{Var}\left( \langle K^{\left(l\right)} \rangle\right)$ and $\langle K^{\left(l\right)}\rangle$ to fit a scaling behaviour $\operatorname{Var}\left(\langle K\rangle\right)\propto \langle K\rangle^{\alpha}$ with $\alpha\in\left[1,2\right]$. Here exponent $\alpha$ serves as an indicator of unit relations, which approaches to $1$ if units become independent and increases to $2$ when units are perfectly correlated.

Meanwhile, we can analyze the scaling of effective free energy. Let us consider $P_{s}\left(\langle K^{\left(l\right)}\rangle\right)$, the probability for a macro-unit to be silent (i.e., exhibits no dynamics) in $X^{\left(l\right)}$, which is equivalent to the probability for all the initial units aggregated into this macro-unit to be silent. This silence probability defines the effective free energy
\begin{align}
    F\left(\langle K^{\left(l\right)}\rangle\right)=-\log P_{s}\left(\langle K^{\left(l\right)}\rangle\right).\label{EQ18}
\end{align}
After obtaining the sequences of $F\left(\langle K^{\left(l\right)}\rangle\right)$ and $\langle K^{\left(l\right)}\rangle$, we can study the scaling behaviour of free energy $F\left(\langle K\rangle\right)\propto \langle K\rangle^{\beta}$. Exponent $\beta\in\left[0,1\right]$ reflects the decay rate of the silence probability. When all units are independent, a fast decrease of the silence probability occurs to imply $\beta=1$. Given perfectly correlated units, we expect a slow reduction characterized by $\beta=0$.

Moreover, we can study the rank scaling of covariance matrix eigenvalue spectrum. For each macro-unit in $X^{\left(l\right)}$, we find all the initial units aggregated into it and calculate a covariance matrix among these initial units (note that this covariance matrix is different from the one used in Eq. (\ref{EQ14}), which is calculated among all macro-units in $X^{\left(l\right)}$). Then, we sort the eigenvalues of this covariance matrix in a decreasing rank. After deriving the ranked eigenvalue spectra of all macro-units in $X^{\left(l\right)}$, we average them to obtain the mean ranked covariance matrix eigenvalue spectrum. Once we obtain the sequence of $\langle K^{\left(l\right)}\rangle$ and mean ranked covariance matrix eigenvalue spectrum, we can fit a rank scaling behaviour
\begin{align}
    \lambda_{r}\propto\left(r/\langle K\rangle\right)^{-\mu},\label{EQ19}
\end{align}
where $\lambda_{r}$ is the $r$-th largest eigenvalue. 

Furthermore, we can explore the dynamic scaling of correlation function. We consider the mean autocorrelation function (i.e., averaged across units) that is maximized at $t=0$
\begin{align}
    C\left(t,\langle K^{\left(l\right)}\rangle\right)=\frac{1}{N^{\left(l\right)}}\sum_{i=1}^{N^{\left(l\right)}}\frac{\Big\langle Y^{\left(l\right)}_{i}\left(0\right)Y^{\left(l\right)}_{i}\left(t\right)\Big\rangle-\left(\mu^{\left(l\right)}_{i}\right)^{2}}{\nu^{\left(l\right)}_{i}-\left(\mu^{\left(l\right)}_{i}\right)^{2}}.\label{DEQ4}
\end{align}
This autocorrelation function exhibits an exponential decay when $t$ departures from $0$, i.e., $C\left(t,\langle K^{\left(l\right)}\rangle\right)=\exp\left(-t/\tau_{c}\left(\langle K^{\left(l\right)}\rangle\right)\right)$. The decay rate is determined by the characteristic time scale, $\tau_{c}\left(\langle K^{\left(l\right)}\rangle\right)$, which can be estimated using the least square fitting. After deriving the sequences of $\tau_{c}\left(\langle K^{\left(l\right)}\rangle\right)$ and $\langle K^{\left(l\right)}\rangle$, we can investigate the potential scale-invariance in system dynamics by verifying whether the re-scaled autocorrelation functions (i.e., let $t\rightarrow t/\tau_{c}\left(\langle K^{\left(l\right)}\rangle\right)$ for each $l$-th iteration) exhibit the universal collapse (i.e., all collapse onto a similar curve). Meanwhile, we can also verify the existence of a scaling behaviour, $\tau_{c}\left(\langle K\rangle\right)\propto \langle K\rangle^{\theta}$. These properties reveal whether system dynamics satisfies the dynamic scaling. 

\subsection{Dataset and pre-processing}
In this section, we present the details of data set preparation.

\paragraph*{\textbf{Cosmic web}} The cosmic web data set is released in Ref. \cite{CosmologicalDataBase}. It is generated by a high resolution cosmological magnetohydrodynamics simulation, which covers a cubic cosmic volume of $100^{3} \text{Mpc}^{3}$ (here $1\text{Mpc}=3.085\times10^{24}\text{cm}$). There are $2400^{3}$ cells and dark matter particles implemented in simulation (see Ref. \cite{vazza2020quantitative} for details). To generate the cosmic web, the cosmic volume is subdivided into four slices (the thickness is $25 \text{Mpc}$). In each slice, the galaxies whose masses are no less than the Milky Way are treated as units in the cosmic web, and the filaments of ordinary and dark matters connecting between galaxy clusters serve as edges. 

\paragraph*{\textbf{Facebook network}} The anonymized Facebook friendship network is offered in Ref. \cite{FacebookDataBase}. It is collected from survey participants in Facebook, covering 	88234 relations among 4039 participants.

\paragraph*{\textbf{Fruit fly central brain network}} The raw data of the fruit fly central brain network is offered by Ref. \cite{xu2020connectome}. It serves as the most fine-grained and large-scale connectome of the fruit fly central brain to date \cite{xu2020connectome}. The connectome spans $\sim 2.5\times 10^{5}$ nm in each dimension, which contains $21662$ traced (i.e., all the branches within the volume are reconstructed) and un-cropped (i.e., main arbors are contained in the volume) neurons as well as $4495$ traced, cropped, and large ($\geq 1000$ synaptic connections) neurons. There exist $\sim 6\times 10^{6}$ traced and un-cropped pre-synaptic sites as well as $\sim 1.8\times 10^{7}$ traced and un-cropped post-synaptic densities. We use the \emph{neu}Print system to acquire this connectome (see Ref. \cite{xu2020connectome} for details) and filter neurons and synapses according to whether their cell bodies are positioned precisely (i.e., assigned with a $10$-nm-spatial-resolution coordinate). The filtered data consists of $23008$ neurons, $4967364$ synaptic connections, and $635761$ pairs of synaptic adjacent relations. We construct the central brain network using synaptic adjacent relations.

\paragraph*{\textbf{Gene co-expression networks}} The gene co-expression networks of cancer phenotypes and healthy bone marrow are released by Ref. \cite{nakamura2023network}, which are estimated from the expression profiles of different gene couples. The considered cancer phenotypes include B and T-cell acute leukemia, acute myeloid leukemia, as well as multiple myeloma. Note that the layouts of these large networks during visualization are derived using the combination of Graph2Vec \cite{narayanan2017graph2vec} and UMAP embedding \cite{mcinnes2018umap}. 

\paragraph*{\textbf{Contact process}} The contact process is generated using the algorithms released by Ref. \cite{nicoletti2020scaling}. A two-dimensional lattice is defined in our experiment, which spans $10^{2}$ in each dimensional and forms $10^{4}$ units. Each unit exhibits dynamics for $5\times 10^{6}$ time steps. The critical condition is realized by setting the spreading rate as $1.649$ while the super-critical condition refers to a case where the spreading rate is $4$.

\paragraph*{\textbf{Whole-brain dynamics}} The whole brain dynamics data of zebrafish larva is offered in Ref. \cite{van2023neural}. There are eight brains in the data set. The activities of $40709\pm 13854$ neurons during $1514\pm 238$ seconds are recorded in each brain, where the time sampling rate is $\simeq 4$.

\subsection{Auxiliary functions for analysis}
In this section, we explain all the auxiliary functions used in our work. These functions are proposed in previous studies and used to support our analysis.

\paragraph*{\textbf{Community detection evaluation}} To evaluate the quality of community detection in Fig. \ref{G3}a, we measure the modularity, separability (i.e., also referred to as the coverage), and correctness (i.e., also referred to as the performance) of communities \cite{fortunato2010community}. In general, the modularity reflects how a network partition formed by communities departs from its null state (i.e., being purely random). The separability counts the fraction of intra-community edges within all edges. The correctness measures the number of correctly interpreted unit pairs (i.e., a pair of units within the same community are expected to connect with each other and two units belonging to different communities should be disconnected).

\paragraph*{\textbf{Unit attributes in networks}} To describe the features of an arbitrary unit in a network in Fig. \ref{G3}c, we measure the local clustering coefficient associated with this unit \cite{saramaki2007generalizations}, degree centrality \cite{borgatti2005centrality}, closeness centrality \cite{freeman2002centrality}, and information centrality \cite{brandes2005centrality}. Among these attributes, the local clustering coefficient counts the fraction of existing triangles associated with a unit within all the possible triangles that may cover this unit. The degree, closeness, and information centrality defines unit centrality in a network according to degrees, the average shortest path distance from one unit to other units, and effective resistance.

\paragraph*{\textbf{Graph Fourier transform and signal gradient}} To analyze graph signals in Fig. \ref{G3}, we apply the PyGSP \cite{pygsp} to realize graph Fourier transformation and gradient measurement. As shown in spectral graph theory \cite{ortega2018graph,dong2020graph}, the frequency spectrum of a graph signal reflects the smoothness of this signal in a network (i.e., how the signal values on adjacent units are similar with each other). As the graph signal becomes smoother, its gradients over the network approach to zero and more energies of the frequency spectrum are concentrated in low frequencies. Because graph signals in our analysis are defined using unit attributes, un-smooth signals suggest the enlarged differences among units.

\paragraph*{\textbf{Clustering evaluation}} To assign the quality of clustering in Fig. \ref{G3}g, we measure the adjusted mutual information \cite{vinh2009information}, completeness \cite{rosenberg2007v}, and adjusted rand score \cite{steinley2004properties} to reflect the consistency between clustering results and their ground-truth references. They are common metrics in clustering performance evaluation.

\subsection{Computational efficiency evaluation}

\paragraph*{\textbf{Computational complexity analysis}}For structure renormalization, the single-step complexity (i.e., the complexity to run one iteration) of the RRG is $O\left(N^{1+\varepsilon}D+M\right)$, where $N$ is the number of units in $X^{\left(1\right)}$, notion $D$, as we defined before, is the dimension of each feature vector in $Y^{\left(1\right)}$, and $M$ is the number of edges (i.e., adjacent relations) \cite{indyk1998approximate,christiani2019fast}. Parameter $\varepsilon\in\left(0,1\right)$ denotes the locality-sensitivity of hashing \cite{indyk1998approximate,christiani2019fast}, which is a function of $h$, the projection dimension defined for hashing in Eq. (\ref{EQ3}), Eq. (\ref{EQ5}), and Eq. (\ref{EQ9}). In general, a larger $\varepsilon$ ensures higher precision of hashing (i.e., Eqs. (\ref{EQ3}-10) hold more robustly), which requires a larger dimension of the signed random projection \cite{indyk1998approximate,christiani2019fast}.

For dynamics renormalization, the complexity of the RRG is $O\left(N^{1+\varepsilon}D\right)$, where $M$ is dropped from complexity measurement since there is no edge to traverse. 

\paragraph*{\textbf{Experiment environment}} All tests in Fig. \ref{G4} are implemented in a 256GB environment with two Intel Xeon Gold 5218 processors. All algorithms are implemented in Python.


 \section*{Acknowledgements}
This project is supported by the Artificial and General Intelligence Research Program of Guo Qiang Research Institute at Tsinghua University (2020GQG1017) as well as the Tsinghua University Initiative Scientific Research Program. Authors appreciate Hedong Hou at the Institut de Math{\'e}matiques d'Orsay and Aohua Cheng at Tsinghua University for their inspiring discussions.

\end{document}